\RequirePackage[2020-09-02]{latexrelease} 

\PassOptionsToPackage{hyphens}{url}
\PassOptionsToPackage{obeyspaces}{url}
\PassOptionsToPackage{spaces}{url}
\PassOptionsToPackage{prologue,dvipsnames}{xcolor}
\PassOptionsToPackage{bookmarksnumbered,unicode}{hyperref}

\documentclass[sigconf]{acmart}

\makeatletter
\def\@ACM@checkaffil{
    \if@ACM@instpresent\else
    \ClassWarningNoLine{\@classname}{No institution present for an affiliation}%
    \fi
    \if@ACM@citypresent\else
    \ClassWarningNoLine{\@classname}{No city present for an affiliation}%
    \fi
    \if@ACM@countrypresent\else
        \ClassWarningNoLine{\@classname}{No country present for an affiliation}%
    \fi
}
\makeatother

\copyrightyear{2022}
\acmYear{2022}
\setcopyright{rightsretained}
\acmConference[CCS '22]{Proceedings of the 2022 ACM SIGSAC Conference on 
Computer and Communications Security}{November 7--11, 2022}{Los Angeles, CA, USA}
\acmBooktitle{Proceedings of the 2022 ACM SIGSAC Conference on Computer and 
Communications Security (CCS '22), November 7--11, 2022, Los Angeles, CA, USA}
\acmDOI{10.1145/3548606.3563517}
\acmISBN{978-1-4503-9450-5/22/11}

\usepackage[utf8]{inputenc}

\usepackage{graphicx}
\usepackage{fancyvrb}
\usepackage{multirow}
\usepackage{balance}
\usepackage[ruled,vlined]{algorithm2e}
\usepackage{array}
\newcolumntype{H}{@{}>{\setbox0=\hbox\bgroup}c<{\egroup}}
\newcommand{\servfail}{{\tt SERVFAIL}}
\usepackage{enumitem}
\setlist{noitemsep}
\usepackage{pifont}

\newcolumntype{P}[1]{>{\centering\arraybackslash}p{#1}}

\usepackage{tikz}
\newcommand*\circled[1]{\tikz[baseline=(char.base)]{
            \node[shape=circle,draw,inner sep=1pt,semithick] (char) {\small #1};}}

\usepackage[printwatermark]{xwatermark}
\newwatermark[allpages,color=black!100,angle=0,scale=1,xpos=0,ypos=-126]{\small CCS '22: Proceedings of the 2022 ACM SIGSAC Conference on Computer and Communications Security, November 2022, Pages 3363–3365 \\
Accepted version. \url{https://doi.org/10.1145/3548606.3563517}}
\usepackage{cleveref}

\begin{document}

\title{Poster: The Unintended Consequences of Algorithm Agility in DNSSEC}

\author{Elias Heftrig}
\affiliation{\small{ATHENE\\Fraunhofer SIT}}

\author{Haya Shulman}
\affiliation{\small{ATHENE\\Fraunhofer SIT \& Goethe-Universität Frankfurt}}

\author{Michael Waidner}
\affiliation{\small{ATHENE\\Fraunhofer SIT \& TU Darmstadt}}



\begin{abstract}
Cryptographic algorithm agility is an important property for DNSSEC: it allows easy deployment of new algorithms if the existing ones are no longer secure. 

In this work we show that the cryptographic agility in DNSSEC, although critical for provisioning DNS with strong cryptography, also introduces a vulnerability.
We find that under certain conditions, when new algorithms are listed in signed DNS responses, the resolvers do not validate DNSSEC. As a result, domains that deploy new ciphers may in fact cause the resolvers not to validate DNSSEC. We exploit this to develop DNSSEC-downgrade attacks and experimentally and ethically evaluate them against popular DNS resolver implementations, public DNS providers, and DNS services used by web clients worldwide. We find that major DNS providers as well as 45\% of DNS resolvers used by web clients are vulnerable to our attacks. 
\end{abstract}
\maketitle

\vspace{-5pt}
\section{Introduction}
DNSSEC [RFC4033-RFC4035] was designed to prevent DNS cache poisoning attacks \cite{kaminsky:dns,cns:frag:dns,dai2021hijackers,brandt2018domain} by authenticating records in DNS responses. Despite costly adoption DNSSEC is slowly gaining traction and increasingly more networks now support DNSSEC. Unfortunately, many domains are signed with algorithms that are no longer considered secure and DNSSEC-supporting DNS resolvers are able to validate only a small number of algorithms. Replacing the existing ciphers or adding new ciphers to DNSSEC is challenging. 

{\bf Cryptographic algorithms agility.} Initially, the DNSSEC standard allowed domains to use either DSA/SHA1 or RSA/SHA1 for signing their zones [RFC4034] - these are no longer deemed secure. Since then, additional algorithms were included in DNSSEC [RFC5155,5702,5933,6605,8080]. The domain owners can now use any subset of 13 algorithms for signing their zones \cite{iana:dnssec:numbers}. In this work our goal is to understand the implications of deployment of new algorithms on the security of DNS. {\em We show that the current state of algorithm agility in DNSSEC introduces a vulnerability, we demonstrate how to exploit it to downgrade DNSSEC validation.} Through analyses of DNSSEC RFCs, public DNS resolvers and DNS implementations we identify the vague recommendations for handling new ciphers to be one of the main factors for the vulnerabilities.

{\bf Unclear specifications for handling unknown ciphers.} According to DNSSEC standard, when returning a lookup result in a signed zone a DNSSEC supporting resolver should either return correctly validated records with an AD flag set, to signal authenticated data, or should return \servfail\ when the data cannot be authenticated. However, the DNSSEC standard does not clearly specify the recommended behaviour for DNS resolvers when faced with new ciphers. Should the resolvers accept records that are signed with unknown algorithms or reject them? How should the validation proceed when a domain supports multiple algorithms? How should the resolvers react in case of inconsistencies in keys between the parent and the child zones? We experimentally show that this lack of clear specification in the DNSSEC standard leads to different vulnerable behaviour implementations at the resolvers: in presence of unknown algorithms in DNSSEC records the resolvers accept the records in the responses without validating them. Even if the signatures are invalid, some resolvers do not return \servfail, but instead accept the DNS records without validation.

{\bf DNSSEC downgrade attacks.} We show how to exploit that resolver behaviour to downgrade DNSSEC validation in resolvers even for zones that are signed with widely known algorithms, such as RSASHA1. The idea behind our attacks is to manipulate the algorithm numbers in DNSSEC records, e.g., DNSKEY, DS, RRSIG. This causes resolvers not to apply DNSSEC validation over DNS records, and exposes them to cache poisoning attacks. 

{\bf Related work on vulnerabilities in DNSSEC.} Previous work \cite{shulman2017one} identified vulnerabilities in keys generation in DNSSEC, which allowed off-path adversaries to compute the secret signing keys of victim domains. This exposed the affected domains to cache poisoning attacks. In this work we exploit vulnerable interpretation of the standard in DNSSEC implementations of DNS resolvers. In 2016 \cite{dai2016dnssec} found that many DNSSEC deployments were misconfigured, e.g., resolvers could not establish a chain of trust to target domains due to expired, missing or inconsistent keys between parent and child domains. In this work we find that only about 0.27\% of popular signed domains have misconfigurations.

{\bf Ethical considerations.} We initiated the notifications of the DNS software vendors and public DNS providers, which we found vulnerable in our work, already in 2021. We experimentally evaluated the attacks reported in this work against servers that we set up as well as against open DNS resolvers and against resolvers of web clients in the Internet using domains that we control. This allowed us to validate the downgrade attacks without downgrading the DNSSEC-security of real domains.

{\bf Contributions.} We experimentally evaluated the attacks reported in this work against servers that we set up as well as against open DNS resolvers and public DNS resolvers, and against resolvers of web clients in the Internet using domains that we control. This allowed us to validate the downgrade attacks without downgrading the DNSSEC-security of real domains. In 2021 we found that some major DNS providers, such as OpenDNS, Google Public DNS and Cloudflare, were vulnerable to our downgrade attacks, and already patched the vulnerabilities. In our Internet wide study, we also find about $13\%$ of open resolvers and almost $45\%$ of the resolvers used by web clients to be vulnerable to downgrade attacks. 
We provide recommendations for preventing our DNSSEC downgrade attacks.
\begin{figure}[t!]
    \centering
    \includegraphics[width=.7\columnwidth]{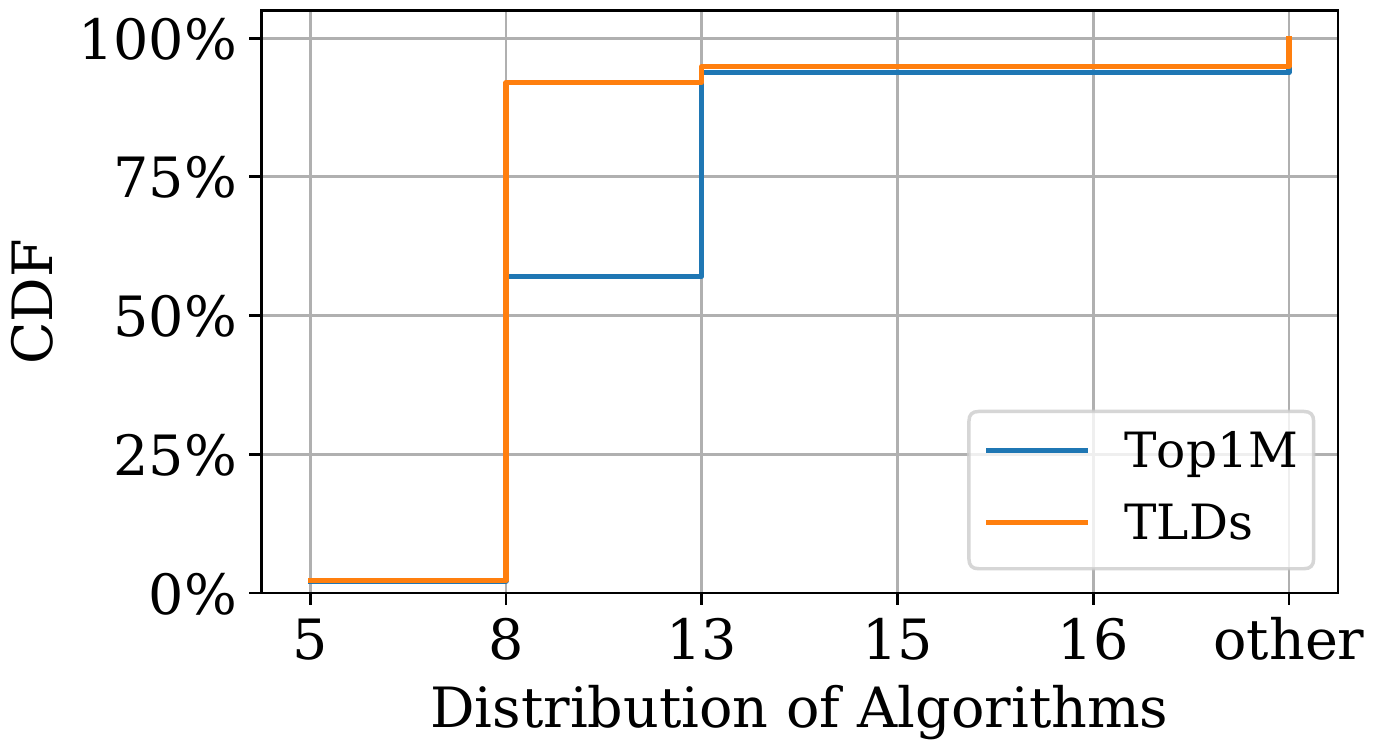}
    \vspace{-10pt}
    \caption{Algorithms in DNSKEY in 1M-Tranco and TLDs.}
    \vspace{-10pt}
    \label{fig:algo_count_merged_cdf_plot}
\end{figure}

\vspace{-5pt}
\section{{\hspace{-3pt}}Factors Exposing to Vulnerabilities}\label{sc:factors}

Our downgrade attacks exploit the vague specification for resolvers' behavior in presence of: (1) unknown algorithms in DNSSEC records, as well as (2) mismatches between the algorithms defined in DS record set (RRset) at the parent and the DNSKEY and the RRSIG records.

In our analysis of resolvers we found that the standard [RFC4035] does not apply in the following configurations: (1) when records are signed with several algorithms, some of which are supported or (2) when the chain of trust cannot be established, e.g., the records are signed with an algorithm that is not present in the DS records.

Another problem is that the DNSSEC standard does not provide details for handling bogus RRsets, and the specific behaviour of resolvers in such cases depends on the choices made by the developers. Some implementations return the unauthenticated records to the calling applications while others return a \servfail\ response code instead. The variations in the interpretation of the standard and choices made by developers and operators indicate the lack of understanding and the lack of consensus on best practices.
\vspace{-5pt}
\section{DNSSEC-Downgrade Attacks}\label{sc:methodology}\label{sc:dataset}\label{sec:attack-methodology}\label{ssc:scenarios} 
In this section we develop methodologies for downgrading DNSSEC validation of resolvers for records in signed domains. Our attacks cause the resolvers to accept fake DNS records with invalid signatures, without validating the signatures.

{\bf Dataset Collection.} Our dataset consists of the following resolvers: (1) popular resolver software implementations, (2) DNS resolvers used by web browsers we targeted with an ad network, (3) open DNS resolvers (including popular free DNS resolver services, such as Google DNS and OpenDNS). 

We find that among the validating DNS resolvers in our dataset, all support the RSA-based DNSSEC algorithms, almost all support the ECDSA algorithms with numbers $13$ and $14$. Most of the DNS resolvers do not support algorithm 16 (ED448) with a gradually increasing support of validation of algorithm 15 (ED25519).

Our dataset of domains contains 1M-top Tranco domains and Top Level Domains (TLDs). DNSSEC is currently deployed for the DNS root zone using algorithm 8 (RSASHA256).
 Out of $1,498$ TLDs, $1,372$ ($91.59\%$) have a DS record at the root zone. In second level domains 43,181 (4.46\%) are DNSSEC signed and have a DS record at the parent. Most domains are signed with a subset of algorithms 5, 8 and 13, and with 5\% using additionally algorithm 15. We plot statistics in Figure \labelcref{fig:algo_count_merged_cdf_plot}.

\begin{figure}[t!]
\vspace{-5pt}
    \centering
    \includegraphics[width=0.38\textwidth]{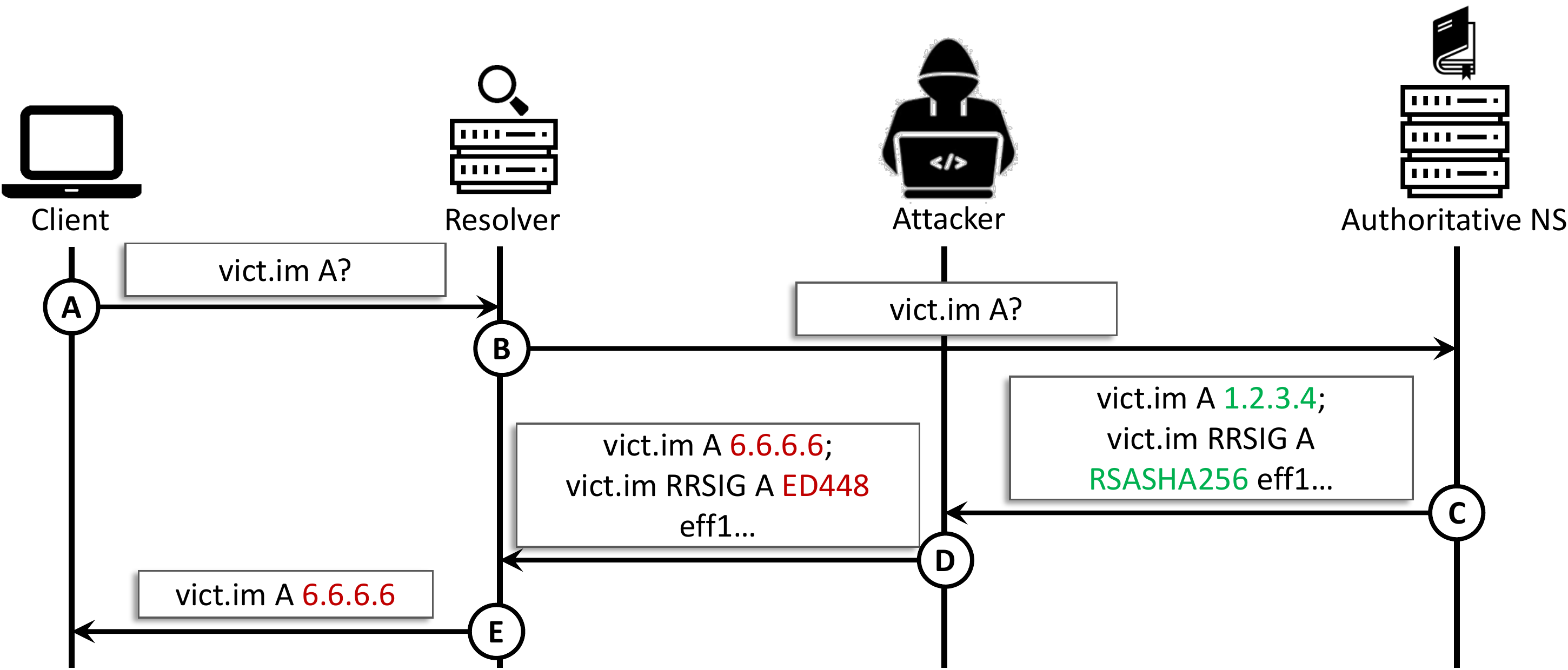}
    \caption{DNSSEC downgrade and injection of fake A record.}
    \vspace{-10pt}
    \label{fig:attack-mitm}
\end{figure}

{\bf Disable validation of DNS responses.} The basic idea behind our downgrade attacks is to manipulate the algorithm number in DNSSEC records to an algorithm number that the resolver does not support. The attack is illustrated in Figure \labelcref{fig:attack-mitm}. Adversary causes the recursive resolver to issue a query. There are different ways to do that, in an example in Figure \labelcref{fig:attack-mitm} we illustrate query triggering using web clients that download our object via an ad network that we deployed. The client sends a query to the recursive resolver (step \circled{A}), which in turn sends it to the authoritative DNS server of the victim domain (step \circled{B}). The response of the nameserver is in step \circled{D} changed by an adversary, who manipulates the algorithm field in the RRSIG record to an unknown number and injects a malicious DNS record. The new record would not pass DNSSEC validation, since the modified record does not match the digest in the signature. However, since the algorithm number in an RRSIG was changed to an unknown algorithm, the resolver does not validate DNSSEC for that answer and caches the malicious records that poison its cache. This disables DNSSEC validation over records in just one DNS response. We next show how to exploit such disabled validation of one DNS response, to inject into resolver's cache a keypair controlled by an adversary, as a result, hijacking a secure delegation for the victim domain, and being able to inject {\em correctly signed bogus records} into the resolver's cache at any later time point.

%
\section{Evaluation}\label{ssc:public}\label{ssc:open:dns}\label{ssc:ad:net}

We evaluated our downgrade attacks against our dataset of resolvers using our own domains. We evaluate our attacks using a Man-in-the-Middle (MitM) adversary model, against which DNSSEC is meant to provide security. We set up a proxy in front of our nameserver, which is our MitM adversary. The proxy manipulates DNSSEC algorithms in responses. 

{\bf Open DNS resolvers.} To evaluate open resolvers we send queries to them for records in our domain. The proxy manipulates algorithm numbers and we evaluate if the records from the responses get accepted. We found the following public DNS providers in our dataset to be vulnerable to our downgrade attacks: Cloudflare, Google Public DNS, OpenDNS and Adguard. They all also exhibit non-RFC compliant validation behaviour. We also found $13.37\%$ open resolvers in our dataset to be vulnerable to DNSSEC downgrade attacks.

{\bf DNS Resolvers via ad network.} When our script is downloaded by the client it iteratively includes resources (img) from test domains, including a non-DNSSEC-signed domain to signal session finish. The web server logs all the requests and delivers the requested resources. A script then analyses the logs to check for vulnerabilities to our DNSSEC downgrade attack. In our ad network study we covered $1385$ Autonomous Systems (ASes) with publicly routable prefixes. From the covered ASes, our server was accessed by $5.79$ clients per AS on average.
Similarly, the ad network study spanned $155$ countries around the globe, which homed $51.70$ clients on average.

We observed different DNS resolvers' behaviour, out of $2476$ validating DNS resolvers that we studied via an ad network with $44.79\%$ being vulnerable: $978$ DNS resolvers were vulnerable to downgrade attack with domains that use DS with algorithms 15 and 16 concurrently - we find that $4$ of these DNS resolvers belong to Comcast; $921$ use Google DNS and are vulnerable; $276$ use Cloudflare and are similarly vulnerable. Our measurement results indicate that a large portion of Internet clients use resolvers that are vulnerable for at least some DS configuration.
For a successful attack it suffices that \emph{any} of the DS configurations in the target zone's ancestors has a configuration vulnerable at the resolver. We found that there were no significant differences in vulnerabilities between the various geolocations. 

\vspace{-12pt}
\section{Recommendations}\label{sec:specupdate}
DNS developers and public DNS providers should support the validation behaviour recommended in the standard. This would prevent the part of our attacks that exploit non standard behaviour, such as those of Google public DNS, OpenDNS, AdGuard and CloudFlare. Nevertheless, adhering to the standard does not solve all the vulnerabilities. Our analysis of the standard shows that lack of clear behaviour specification for bogus validation outcome introduces a vulnerability which we exploit in our attacks.

As a systematic solution to prevent downgrade attacks we propose to extend the DNSSEC standard to consider situations in which adversaries can turn off DNSSEC protection, by imposing a more rigorous requirement for \servfail\ return codes when an RRset is classified as ``bogus''. This would prevent the attacks without imposing restrictions on the usage of DNSSEC and without limiting the flexibility of deployment of new algorithms. We next describe the recommendations for validation of DNSSEC supporting resolvers.

To validate DNSSEC, a DNSSEC supporting resolver is required to validate all the signature (RRSIG) records over DS RRset. If validation fails the resolver must return \servfail. If none of the algorithms in the DS are supported by the resolver, DNSSEC validation is not applied and the resolver considers the zone as insecure. If the resolver supports the cryptographic algorithm and the digest in at least one DS record, the resolver is required to check that there is a matching DNSKEY RRset in the child zone. Then the resolver uses a signature, that corresponds to the key digest in the supported DS record, to validate the DNSKEY record of the child. If the validation is successful, the chain of trust established and the DNSKEY RRset can be used to validate signatures over records in that zone. If the signature is invalid or if the DNSKEY cannot be found, the resolver should return \servfail\ to any query for that zone. This behaviour would prevent our attacks. 

\vspace{-5pt}
\section{Conclusions}\label{sc:conc}
Cryptographic algorithm agility in DNSSEC, i.e., the ability to add and remove algorithms, is an important requirement needed to maintain strong security guarantees. Algorithms may be broken, being able to replace vulnerable algorithms with secure ones efficiently and fast is critical \cite{housley2015guidelines}. We show that efficient and fast adoption of algorithms also introduces a challenge: how should resolvers react when faced with records signed using new algorithms? What is the correct behaviour with zones that are signed with a number of algorithms, only some of which are known? 

The standard does not provide clear recommendations for resolvers how to handle DNSSEC records with unknown algorithms and how to handle bogus data, but leaves it open for every resolver to make its own decision how to behave in such cases.
We discover that the vague specification leads to different validation behaviour in popular DNS resolver implementations, which indicates that there is no consensus on what a correct behaviour should be. We show that often the resolver behaviour is vulnerable and demonstrate DNSSEC downgrade attacks.
\section*{Acknowledgements}
This work has been co-funded by the German Federal Ministry of Education and Research and the Hessen State Ministry for Higher Education, Research and Arts within their joint support of the National Research Center for Applied Cybersecurity ATHENE and by the Deutsche Forschungsgemeinschaft (DFG, German Research Foundation) SFB~1119.

\bibliographystyle{ACM-Reference-Format}
\bibliography{ref,bib,sec}

\end{document}